\documentclass[iop]{emulateapj}
  
\usepackage{apjfonts}
\usepackage{amsmath}
\usepackage{epsf}
\usepackage{color}
\usepackage{comment}
\usepackage{amsmath}
\usepackage{graphicx}
\usepackage{subfigure}
\usepackage{epstopdf}
\usepackage{float}
\shortauthors{Chiang, Overzier \& Gebhardt}

\begin{document}
\title{Discovery of a large number of candidate proto-clusters traced by $\sim15$ Mpc-scale galaxy overdensities in COSMOS}
%\title{Candidate proto-clusters traced by 15 Mpc scales galaxy distribution in COSMOS}
%\title{Candidate proto-clusters in COSMOS found on scales of 15 Mpc}
%\title{High-Redshift Proto-Clusters Traced by the Large-Scale Galaxy Distribution in COSMOS}
%\title{Systematic Search for High-redshift Cluster Progenitors Using $\sim$10 Mpc Scales Galaxy Distributions: Discovery of Numerous Proto-cluster Candidates in COSMOS}
%\title{Large-scale Galaxy Distribution Traces Forming Galaxy Clusters in COSMOS/UltraVISTA}
%\title{A Large Sample of Candidate Proto-Clusters in COSMOS/UltraVISTA}

\author{Yi-Kuan Chiang\altaffilmark{1}, Roderik Overzier\altaffilmark{2, 1}, Karl Gebhardt\altaffilmark{1}}

\altaffiltext{1}{Department of Astronomy, University of Texas at Austin, 1 University Station C1400, Austin, TX 78712, USA}
\altaffiltext{2}{Observat\'orio Nacional, Rua Jos\'e Cristino, 77. CEP 20921-400, S\~ao Crist\'ov\~ao, Rio de Janeiro-RJ, Brazil}

\begin{abstract}
To demonstrate the feasibility of studying the epoch of massive galaxy cluster formation in a more systematic manner using current and future galaxy surveys, we report the discovery of a large sample of proto-cluster candidates in the 1.62 deg$^2$ COSMOS/UltraVISTA field traced by optical/IR selected galaxies using photometric redshifts. By comparing properly smoothed 3D galaxy density maps of the observations and a set of matched simulations incorporating the dominant observational effects (galaxy selection and photometric redshift uncertainties), we first confirm that the observed $\sim15$ comoving Mpc scale galaxy clustering is consistent with $\Lambda$CDM models. Using further the relation between high-z overdensity and the present day cluster mass calibrated in these matched simulations, we found 36 candidate structures at $1.6<z<3.1$, showing overdensities consistent with the progenitors of M$_{z=0} \sim10^{15}$ M$_{\odot}$ clusters. Taking into account the significant upward scattering of lower mass structures, the probabilities for the candidates to have at least M$_{z=0}\sim10^{14}$ M$_{\odot}$ are $\sim70\%$. For each structure, about $15\%-40\%$ of photometric galaxy candidates are expected to be true proto-cluster members that will merge into a cluster-scale halo by $z=0$. With solely photometric redshifts, we successfully rediscover two spectroscopically confirmed structures in this field, suggesting that our algorithm is robust. This work generates a large sample of uniformly-selected proto-cluster candidates, providing rich targets for spectroscopic follow-up and subsequent studies of cluster formation. Meanwhile, it demonstrates the potential for probing early cluster formation with upcoming redshift surveys such as the Hobby-Eberly Telescope Dark Energy Experiment and the Subaru Prime Focus Spectrograph survey.

\end{abstract}

\section{Introduction}

Galaxy clusters are extreme products of structure formation. They are ideal laboratories to study galaxy assembly, quenching, and sub/super-halo galaxy environments. It has become clear that to leverage a complete picture of cluster formation, we need to find and study also their progenitors at high redshifts that were still forming. In the past decade, a limited number of observations of ``proto-clusters'' revealed some intriguing properties such as sped-up galaxy evolution \citep{steidel05}, abnormal metallicities \citep{kulas13}, and the enhancement of star-forming galaxies \citep{overzier08, hayashi12}, extreme starbursts \citep{capak11}, extended Ly$\alpha$ blobs \citep{matsuda12}, and AGN \citep{lehmer09, martini13}. However, exactly how these clues are related to the formation of clusters as a whole is not yet understood.
%that formed from initial perturbations at very high redshift

Due to the low number density of cluster progenitors and the difficulties in identifying them, only $\sim10$ proto-clusters have been discovered in ``random'' fields \citep[e.g.,][]{steidel98, steidel05, ouchi05, toshikawa12}. Highly biased tracers like radio galaxies and quasars have been used to narrow down the search volume, generating another $\sim10$ structures \citep[e.g.,][]{pentericci00, kurk00, kurk04a, venemans02, venemans04, venemans07, galametz10, trainor12}. However, the number of cluster progenitors not traced by radio galaxies should far exceed the number that does, based on AGN duty cycle arguments \citep{west94}. A recent compilation of structures observed to date can be found in \cite{chiang13}.
%since most of the serendipitously found structures do not associate with radio galaxies.} 
%the number of cluster progenitors not traced by radio galaxies should far exceed the number that does, simply based on AGN duty cycle arguments. 

In \cite{chiang13} we presented the physical properties and observational signatures of proto-clusters predicted in $\Lambda$CDM models using a large set of simulated clusters drawn from the Millennium Run simulations in $WMAP1$ and $WMAP7$ cosmologies. The progenitor regions of galaxy clusters can already be identified at very high redshifts given their significant large-scale (few tens of comoving Mpc) density contrasts compared to the field. Systematic searches in future large galaxy redshift surveys are thus very promising.

Although suffering from significant redshift uncertainties, the current generation of deep and wide photometric surveys may already provide the first large and relatively unbiased sample of cluster progenitors at $z\gtrsim2$. In this Letter, we extend our methods from \cite{chiang13} to the regime of moderate redshift precision to search for cluster progenitors in the 1.62 deg$^2$ COSMOS/UltraVISTA survey using photometric redshifts. The COSMOS field is being targeted by a great number of surveys, many of which are aimed at studying large-scale structure at high redshift \citep{silverman09, kovac10, diener13, kashino13}. Focusing on the correlation between galaxy properties and their environments, \cite{scoville13} have generated a set of galaxy density maps of this field with dynamically varying scales. For the purpose of searching for cluster progenitors, specifically, we generate an alternative set of large-scale galaxy density maps designed to maximize the contrast between cluster progenitors and the field. By comparing the data with a set of matched simulated fields incorporating galaxy selection effects and redshift uncertainties, we have identified a large sample of candidate high-redshift cluster progenitors. Our technique recovers two previously known ``proto-clusters'' (Spitler et al. 2012; Chiang et al. in prep.), suggesting that the algorithm is robust. We present the positions and redshifts of the candidates, together with estimates of their present day descendant masses. We adopt a cosmology [$h$, $\Omega_m$, $\Omega_{\Lambda}$, $n_s$, $\sigma_8]=[$0.73, 0.25, 0.75, 1, 0.9], but note that our results are relatively insensitive to the assumed values of $\sigma_8$ and $H_0$ (see \cite{chiang13}).

\section{Data and Cluster Finding Technique}

The \cite{muzzin13} galaxy catalog covers 1.62 deg$^2$ of the COSMOS/UltraVISTA field with 30 photometric bands from ultraviolet to infrared. The catalog is complete (90\%) to K$_{s}\mathrm{,AB}=23.4$ mag. Photometric redshifts ($z_{phot}$) were computed using the code of \cite{brammer08}. \cite{Ilbert13} compiled an alternative COSMOS catalog (not directly used in this Letter). By comparing with mainly the zCOSMOS-bright and faint spectroscopy  \citep[][in prep.]{lilly07, lilly09}, \cite{scoville13} present the estimated $z_{phot}$ uncertainty of the later catalog as a function of redshift and K$_{s}$ magnitude, where we obtain an estimated average $z_{phot}$ uncertainty of $\sigma_z=0.025(1+z)$ for the galaxies we will be using (K$_{s}<23.4$; $1.5\lesssim z\lesssim3$). In our analysis we will assume that this $\sigma_z$ is the typical uncertainty in the distribution of the galaxies in redshift space. We will show in Section 3 that this value of $\sigma_z$ indeed gives consistent overdensity distributions when comparing the observations with our simulation.

%at  $1.5\lesssim z\lesssim4$
%$\sigma_z=0.03(1+z)$ for a fainter K$_s$ band cut (K$_{s}<24$ AB).
%The $z_{phot}$ have an estimated accuracy of $\sigma_z=0.013(1+z)$ up to $z\sim1.5$ based on the zCOSMOS-bright \citep{lilly07, lilly09} spectroscopy. \cite{Ilbert13} compiled an alternative COSMOS catalog (not directly used in this Letter). By comparing with the zCOSMOS-faint spectroscopy (Lilly et al. in prep.), the estimated photometric redshift uncertainty at  $1.5\lesssim z\lesssim4$ is $\sigma_z=0.03(1+z)$ for a fainter K$_s$ band cut (K$_{s}<24$ AB).

Cluster progenitors manifest themselves by having high galaxy overdensities defined as $\delta_{gal}(\vec{x})\equiv(n_{gal}(\vec{x})-\langle n_{gal}\rangle)/\langle n_{gal}\rangle$, where $n_{gal}(\vec{x})$ is the local galaxy number density in a designated window (specified later) and $\langle n_{gal}\rangle$ is the mean galaxy number density over the whole field. To calculate $\delta_{gal}(\vec{x})$, we use galaxies in the \cite{muzzin13} catalog with K$_{s}\mathrm{,AB}<23.4$. A small fraction ($\sim4\%$) of the galaxies with a broad and/or multi-modal $z_{phot}$ distribution are excluded, but we note that our final results are nearly the same with and without this quality cut.

%The average $z_{phot}$ error in the redshift range of interest ($2\lesssim z\lesssim3$) is likely to be significantly larger than the $0.013(1+z)$ of \cite{muzzin13} since spectral features like the 4000\AA\ and Balmer breaks shift out of the medium-bands at $z\gtrsim1.5$, but smaller than the $0.03(1+z)$ of \cite{Ilbert13} since we use a brighter K$_s$ band cut. This is also demonstrated in \cite{scoville13}. We therefore adopt a value of $\sigma_z=0.025(1+z)$ for further analysis, and we will show in Section 3 that this value indeed gives consistent $\delta_{gal}$ distributions in the observation and simulation.}
%To generate proto-cluster scale overdensity maps,
%and a well localized $z_{phot}$ with >80\% probability that $|z_{true}-z_{phot}|<0.2$.
%the galaxy overdensity $\delta_{gal}(\vec{x})\equiv(n_{gal}(\vec{x})-\langle n_{gal}\rangle)/\langle n_{gal}\rangle$ allows us to distinguish them from the field and predict their present day descendant mass M$_{z=0}$ \citep{chiang13}.}

With these galaxies, we generate a three-dimensional overdensity map of the COSMOS field on a regularly spaced grid with a spacing of 1 arcmin on the sky and 0.01 in redshift. For each grid point, we calculate $\delta_{gal}$ in a cylindrical window with a radius $r=5$ arcmin ($\sim15$ comoving Mpc in diameter at $z\sim2$) and a redshift depth full width of $l_z=\sigma_z=0.025(1+z)$. This window is designed to maximize the contrast between cluster progenitors and field based on the projected size of proto-clusters ($\sim10-30$ comoving Mpc at $z\sim2$) while not over-resolving the galaxy distribution in redshift due to the $z_{phot}$ uncertainties\footnote{Varying the $l_z$ by a factor of $<2$ gives different $1+\delta_{gal}$ by $<10\%$ and our final interpretation of overdense regions stays nearly the same if the same $l_z$ is used to calculate $\delta_{gal}$ in the simulation.}.

To quantify the relation between $\delta_{gal}(z)$, the overdensity, and M$_{z=0}$, the cluster mass at $z=0$, we use a set of simulated observations matched to the COSMOS data set. We start with the 24 $1.4\times1.4$ deg$^2$ lightcones from \cite{henriques12} that are based on the Millennium Run simulations \citep{springel05} and the \cite{guo11} semi-analytic model with the \cite{bc03} stellar population synthesis models. To match our selection of galaxies in COSMOS, we set the same K$_s$ magnitude limit. Next, we implement the $z_{phot}$ errors in a set of Monte Carlo realizations (20 realizations for each of the 24 lightcones). We shuffle the redshifts (which include the line-of-sight peculiar velocity component) of the galaxies in the simulation according to a Gaussian distribution with a $\sigma_z$ similar to that of the real COSMOS sample. Since the simulated catalog contains about twice as many galaxies as the real catalog for K$_{s}<23.4$, we further take out a random subset of the simulated galaxies in these realizations to re-create the same level of Poisson counting errors. Next, we generate $\delta_{gal}$ maps for each realization using the same procedure used for the real data. Because we know the locations of all clusters in the simulations, we can now calibrate the M$_{z=0}-\delta_{gal}(z)$ relation. We extract $\delta_{gal}$ distributions for the whole volume and for clusters in 3 mass bins (``Fornax-'' type: M$_{z=0}=1-3\times10^{14}$ M$_{\odot}$, ``Virgo-'' type: M$_{z=0}=3-10\times10^{14}$ M$_{\odot}$ and ``Coma-'' type: M$_{z=0}>10^{15}$ M$_{\odot}$), allowing us to characterize regions in the COSMOS field according to their overdensities.

%Because the M$_{z=0}-\delta_{gal}(z)$ relation depends highly on the $z_{phot}$ errors

\section{Results}

\subsection{The Large-scale Density Field in COSMOS/UltraVISTA}

\begin{figure*}
\epsscale{1.16}
\plotone{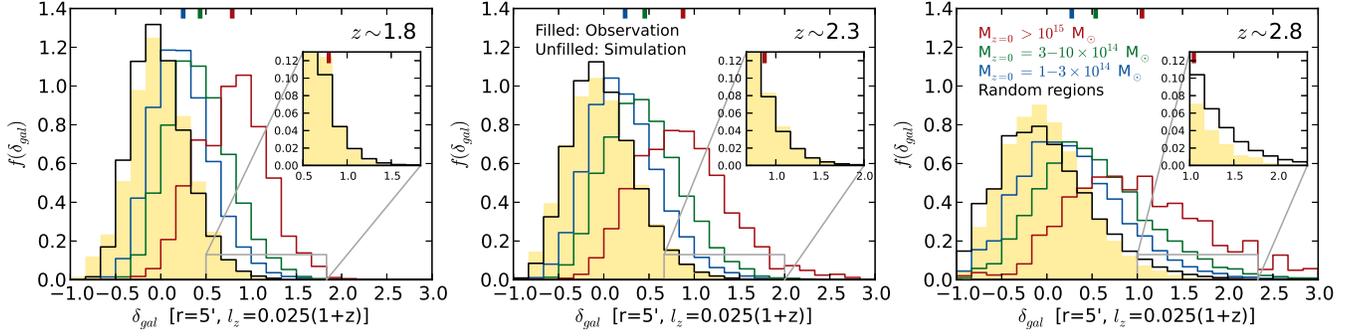}
%\vspace{2mm}
\caption{Probability distribution function (PDF, normalized) of galaxy overdensity for the COSMOS data (filled) and simulated observation (black) at 3 redshift ranges extracted from the whole survey/simulation volumes containing voids, fields, and proto-clusters. The PDFs for the proto-cluster regions in the simulation are shown in blue (``Fornax''), green (``Virgo''), and red (``Coma'') for 3 cluster mass bins, respectively. Thick ticks indicate the mean for each simulated proto-cluster PDF. The $\delta_{gal}$ is calculated in a cylindrical window with $r=5'$ and $0.025(1+z)$ depth in redshift.}
\end{figure*}

\begin{figure*}[t!]
     \begin{center}
        \hspace*{-0.2em}
        \subfigure{%
            \includegraphics[width=0.333\textwidth]{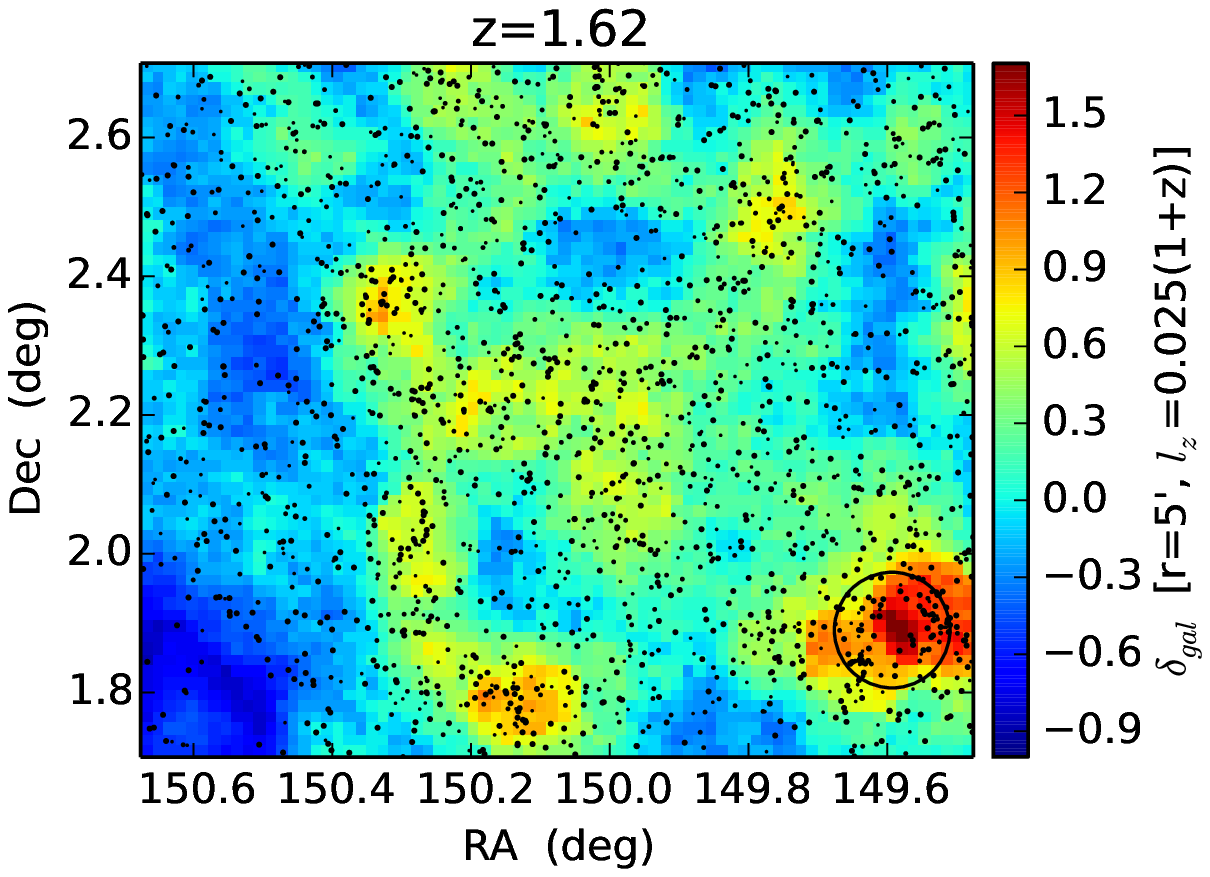}
        }%
        \hspace*{-0.2em}
        \subfigure{%
            \includegraphics[width=0.333\textwidth]{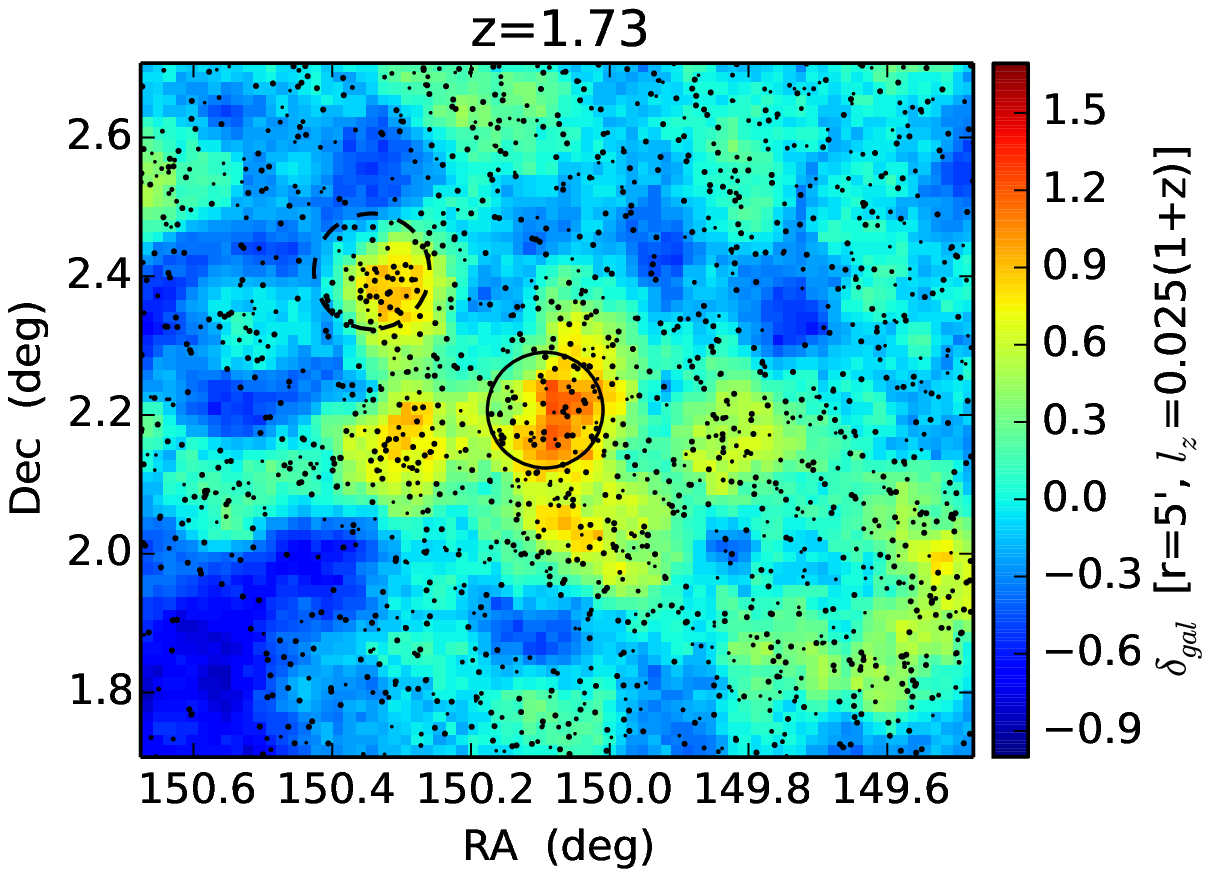}
        }%
        \hspace*{-0.2em}
        \subfigure{%
           \includegraphics[width=0.333\textwidth]{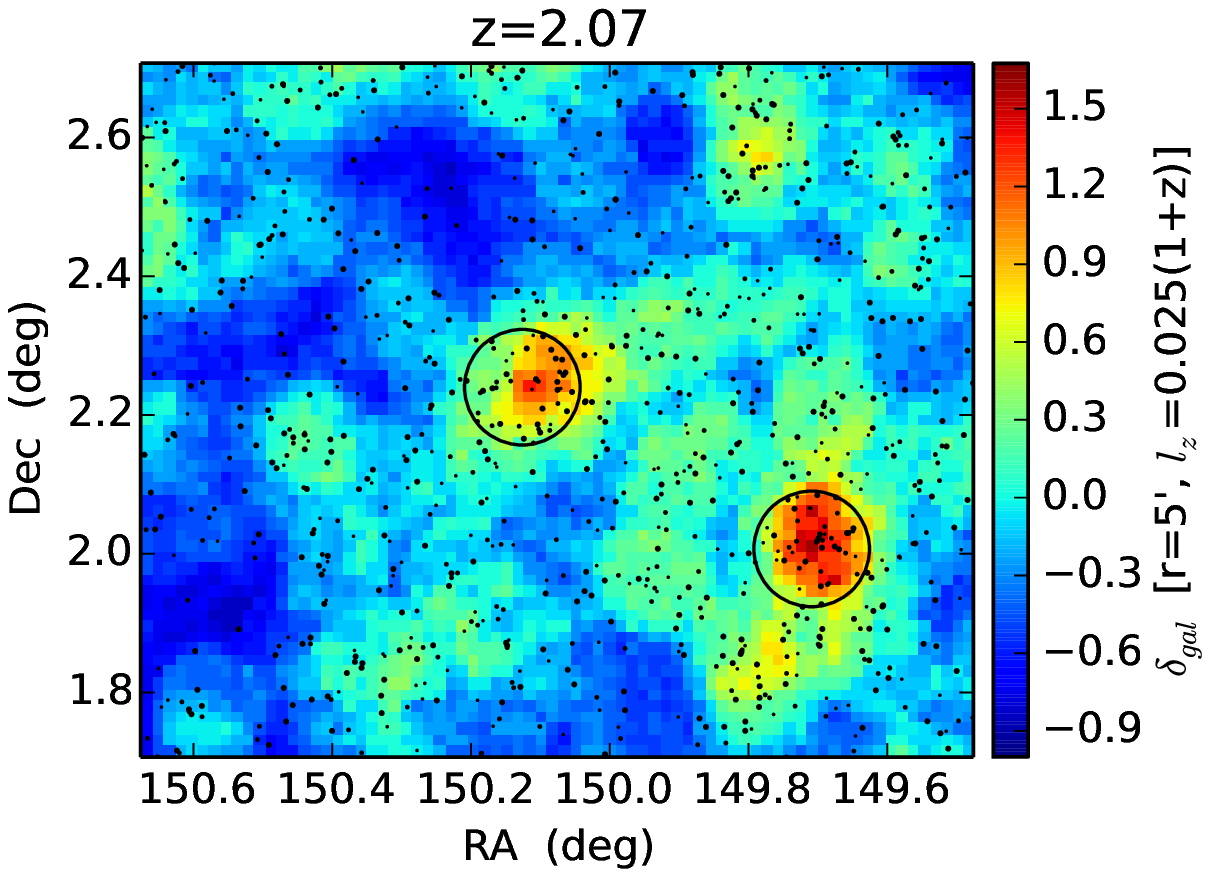}
        }\\ %  ------- End of the first row ----------------------%
        \vspace*{-0.6em}
        \hspace*{-0.2em}
        \subfigure{%
            \includegraphics[width=0.333\textwidth]{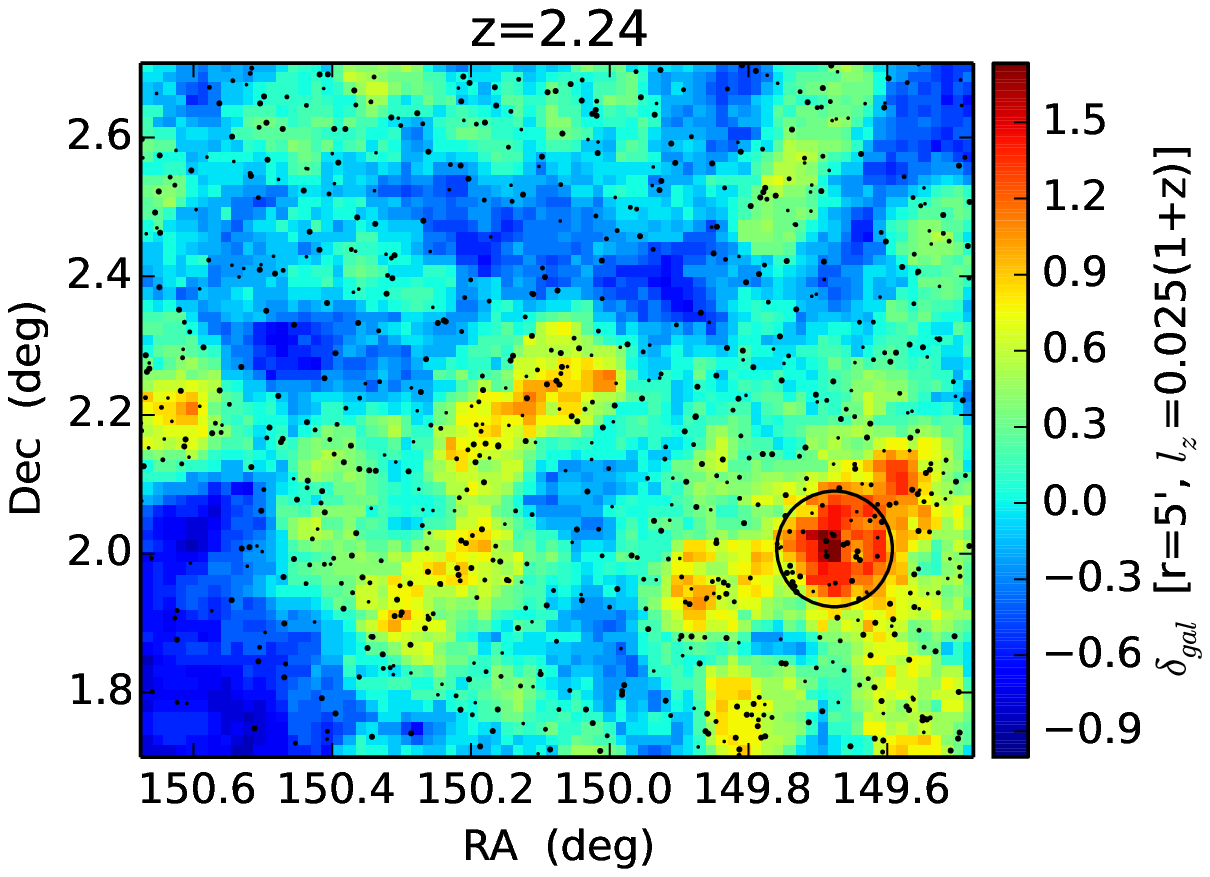}
        }%
        \hspace*{-0.2em}
        \subfigure{%
            \includegraphics[width=0.333\textwidth]{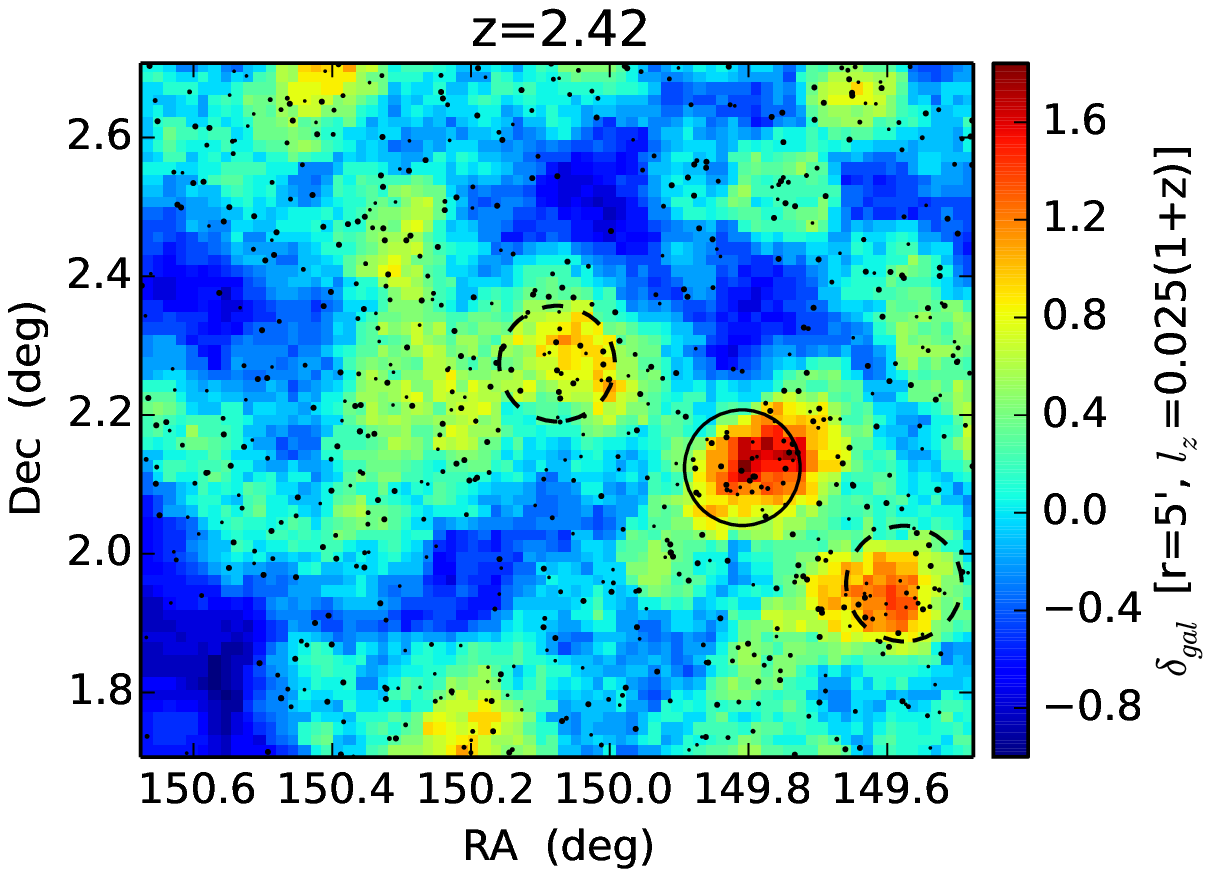}
        }%
        \hspace*{-0.2em}
        \subfigure{%
           \includegraphics[width=0.333\textwidth]{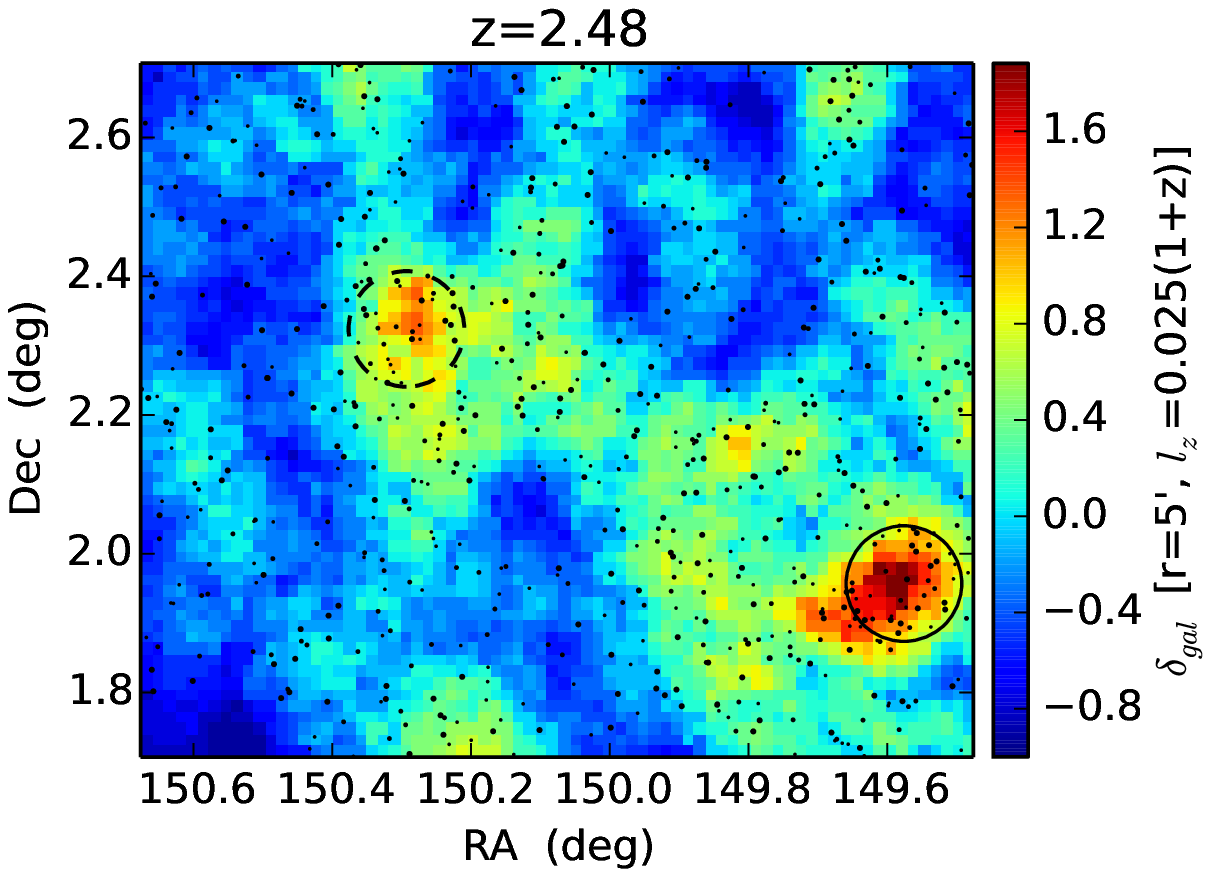}
        }\\ %  ------- End of the Second row ----------------------%
        \vspace*{-0.6em}
        \hspace*{-0.2em}
        \subfigure{%
            \includegraphics[width=0.333\textwidth]{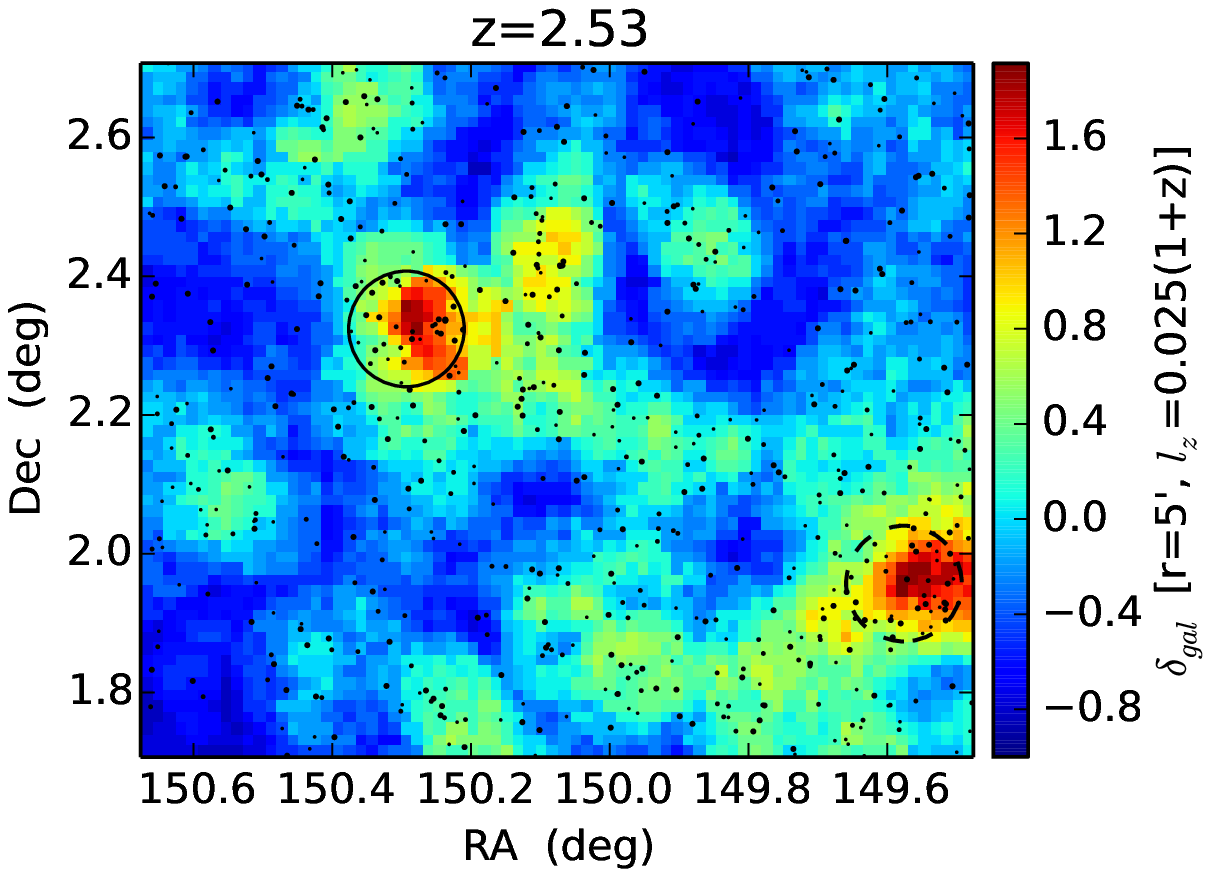}
        }%
        \hspace*{-0.2em}
        \subfigure{%
            \includegraphics[width=0.333\textwidth]{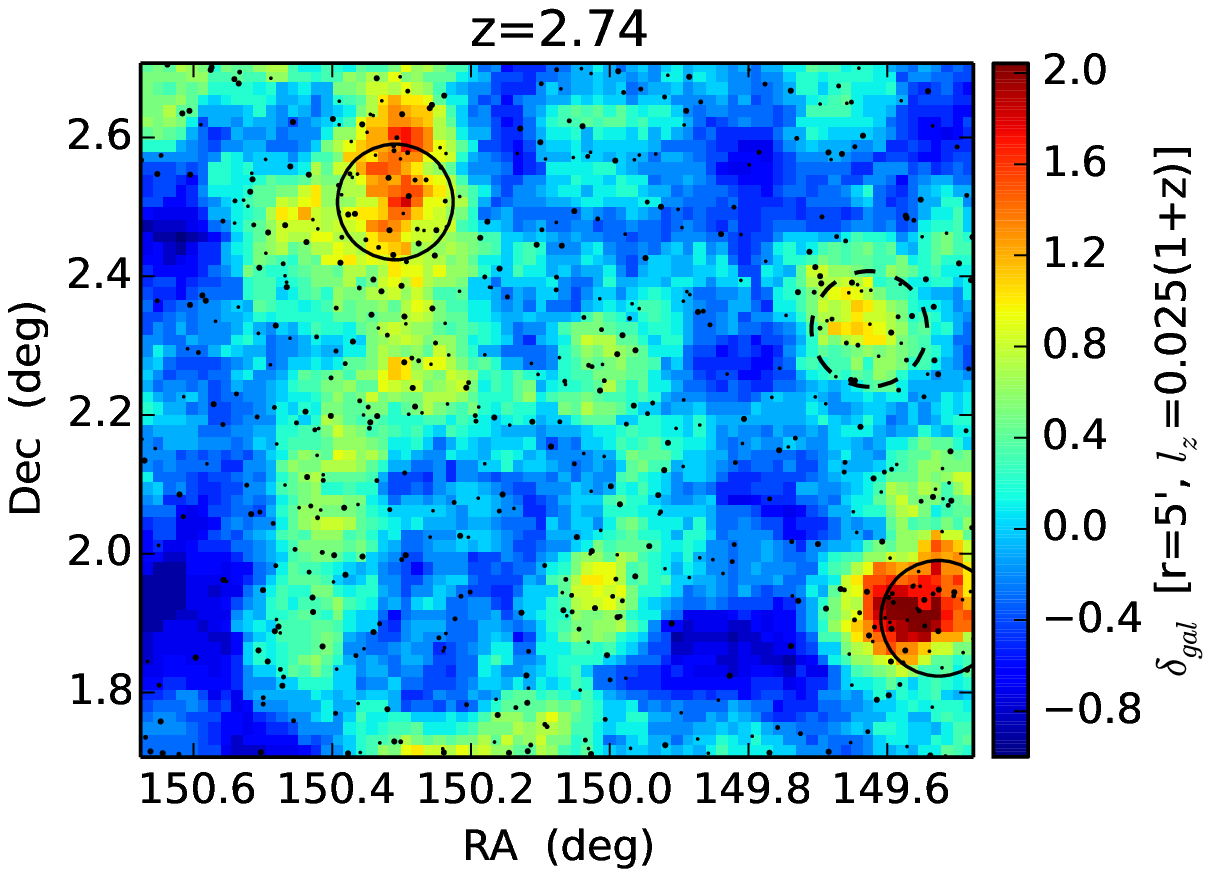}
        }%
        \hspace*{-0.2em}
        \subfigure{%
           \includegraphics[width=0.333\textwidth]{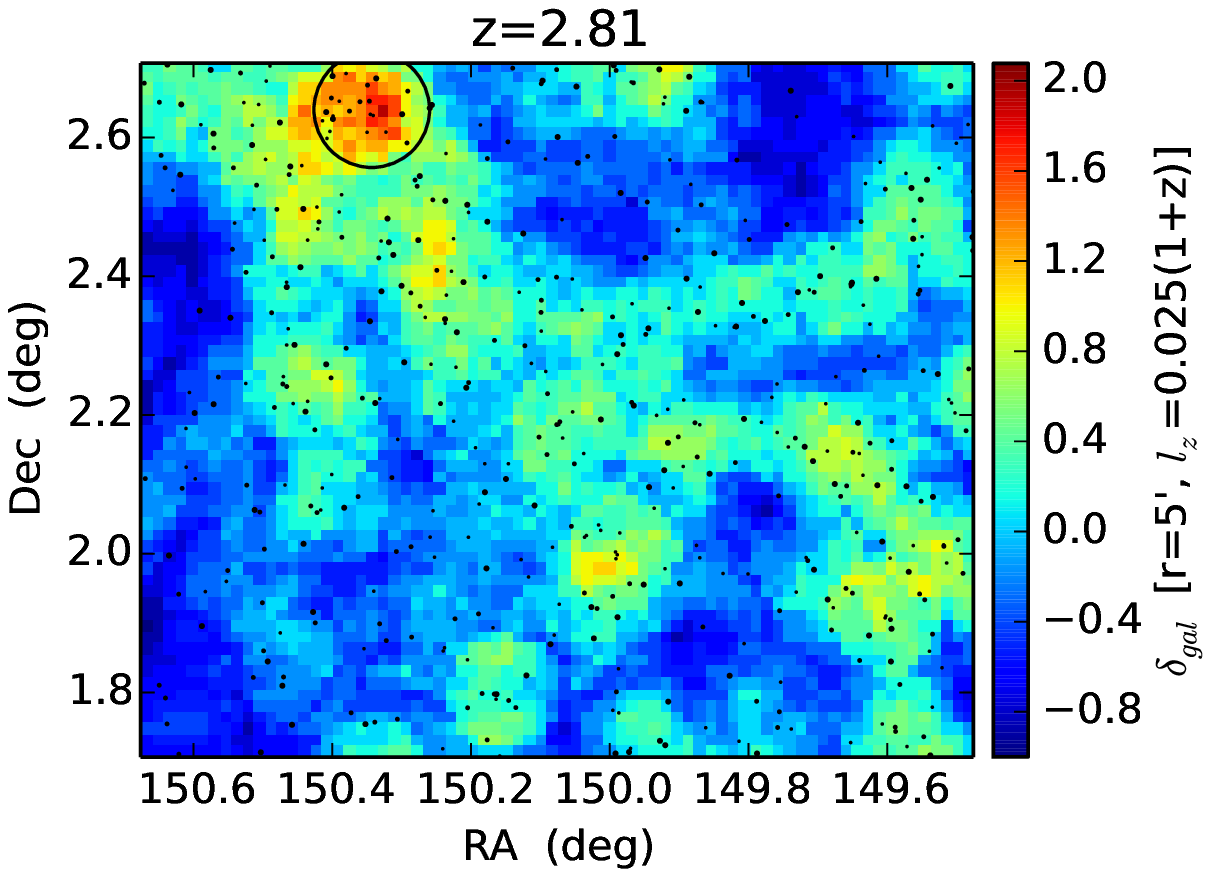}
        }\\ %  ------- End of the third row ----------------------%

        \vspace*{-0.6em}
    \end{center}
    \caption{%
        $\delta_{gal}$ maps for a selection of the most prominent proto-cluster candidates in the COSMOS/UltraVISTA field. Galaxies in the $0.025(1+z)$ redshift depth are marked by dots with sizes scaled by K$_s$ band flux. The color is scaled to turn red for regions more overdense than the average $\delta_{gal}$ of ``Coma-type'' proto-clusters (M$_{z=0}>10^{15}$ M$_{\odot}$) found in the simulation (see Figure 1). The solid and dashed circles indicate the positions of proto-cluster candidates with $\delta_{gal}$ peak at the redshifts shown and adjacent redshifts, respectively.
     }%
   \label{fig:subfigures}
\end{figure*}

\begin{figure*}
\epsscale{1.16}
\plotone{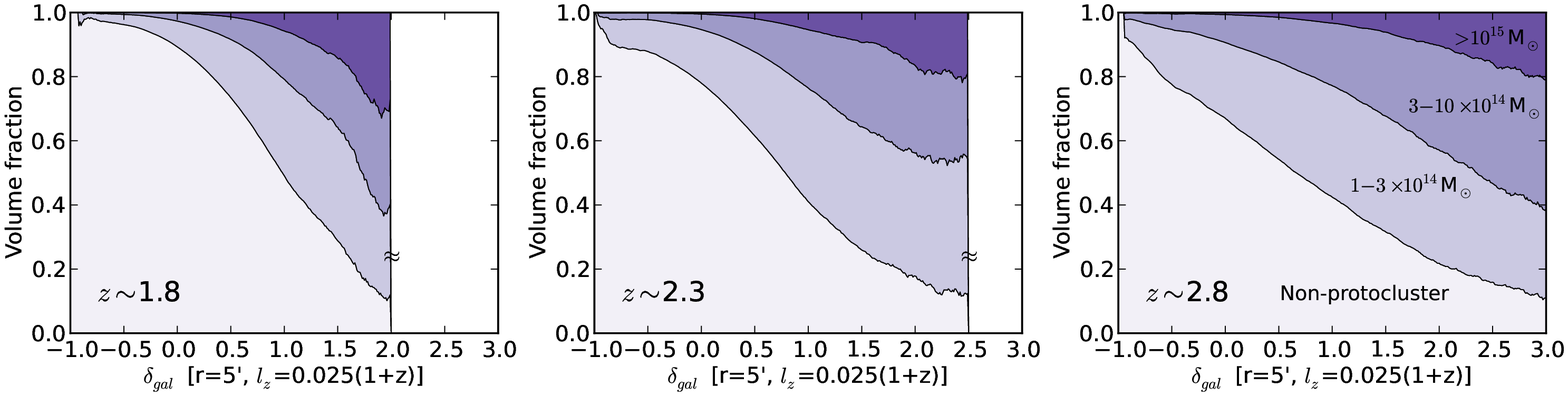}
%\vspace{2mm}
\caption{Volume fractions (probabilities) for a region with a given galaxy overdensity to be non-protocluster (light purple) and proto-cluster with 3 mass ranges (dark purple) at 3 redshift ranges based on the simulation.}
\end{figure*}

Figure 1 shows, with filled histograms (yellow), the probability distributions of $\delta_{gal}$ for COSMOS at 3 redshifts ($z=1.8$, 2.3, and 2.8), with $\delta_{gal}$ defined as in Section 2. Only data in the central $1.2\times1.0$ deg$^2$ region of the field are used to avoid the incompleteness at the edges. The shape of the $\delta_{gal}$ distribution is close to a Gaussian with a high $\delta_{gal}$ tail. Even with the large smoothing scale of $\sim15$ comoving Mpc and moderate redshift projections, the high tails of $\delta_{gal}$ (thus, the departure from Gaussian distribution) are clearly seen, suggesting that structure growth at this scale has evolved toward the non-linear regime expected for forming galaxy clusters.
%\textbf{Compared with low redshift $\delta_{gal}$ distribution,} the more prominent high $\delta_{gal}$ tails at higher redshift mainly reflect a higher bias of the galaxy tracers when using a flux limited sample.

To compare the real data with our matched simulated observations, we overplot in Figure 1 the $\delta_{gal}$ distributions of the simulation (whole volume, including fields and clusters). The $\delta_{gal}$ distributions of the observation (yellow) and simulation (black) match very well\footnote{The differences between the observed and simulated histograms give a reduced $\chi^2$ of $\lesssim1$ for $z\sim1.8$ and $2.3$ and of 1.8 for $z\sim2.8$.}, indicating that the overall large-scale galaxy clustering probed by the COSMOS data is consistent with our $\Lambda$CDM models. This also suggests that our matched simulation re-creates the main observational effects and bias successfully. The $\delta_{gal}$ distribution of cluster progenitors in the simulation are shown as unfilled histograms in Figure 1 for 3 present day cluster mass bins. The mean $\delta_{gal}$ for each mass bin is $\sim0.25$, $\sim0.5$, and $\sim0.8-1.0$ for ``Fornax'', ``Virgo'' and ``Coma'' proto-clusters respectively. The center of the cluster progenitors in the simulation is defined by the K$_s$ band flux weighted average positions of the member galaxies. In general, the progenitors of higher mass clusters show higher $\delta_{gal}$ but there is a certain degree of overlap due to mainly the redshift uncertainties and sub-dominant effects such as the intrinsic scatters of the M$_{z=0}-\delta_{gal}(z)$ relation and Poisson errors.
%compared to the ensemble

To further examine how well our matched simulation fits the data, we show zoom-in regions of the high $\delta_{gal}$ tails in Figure 1. The high $\delta_{gal}$ tails of the simulation and observation match remarkably well especially for the $z\sim1.8$ and 2.3 bins. This indicates that the simulation is a good approximation of the observation and that our interpretation of the observed overdensities will be nearly unbiased\footnote{At $z\sim2.8$, the simulation shows slightly stronger clustering at the high $\delta_{gal}$ end, indicating slightly larger $z_{phot}$ uncertainties for the real data at higher redshifts. This will give us slightly more conservative results (e.g., lower probability to be a cluster progenitor for a given observed $\delta_{gal}$) when interpreting observed structures at $z\gtrsim2.8$.}.

We present, in Figure 2, the $\delta_{gal}$ maps for a selection of redshift slices in the COSMOS field. Each slice has a redshift width of $0.025(1+z)$ and shows a variety of cosmic structures from voids (blue), filaments (green), to overdense peaks (red). The regions in red have a $\delta_{gal}$ well in the high tail regime of the $\delta_{gal}$ distribution (Figure 1) where we expect to find cluster progenitors. We note that our maps are consistent with \cite{scoville13} but differ in the fact that we tend to preserve the large proto-cluster scale overdensities instead of breaking them up into multiple sub-structures.

\begin{table}[t!]
\caption{\label{tab:candidates} Proto-cluster Candidates}
\begin{center}
\begin{tabular}{ccccccccc}
\hline\hline
ID & $z$ & RA & Dec & $\delta_{gal}$$^a$ & $P_{pc}$$^b$ & $P_{Coma}$$^c$ & $z_{2nd}$$^d$ &  \\
\hline

1 & 1.62 & 149.593 & 1.89 & 1.83$^{+0.32}_{-0.29}$ & 0.66 & 0.05 &  & \\
2 & 1.73 & 150.093 & 2.207 & 1.23$^{+0.29}_{-0.25}$ & 0.59 & 0.08 & 1.78 & \\
3 & 1.74 & 150.343 & 2.407 & 1.1$^{+0.27}_{-0.24}$ & 0.54 & 0.06 &  & \\
4 & 1.87 & 149.893 & 1.907 & 1.27$^{+0.31}_{-0.27}$ & 0.66 & 0.12 &  & \\
5 & 1.94 & 150.043 & 2.174 & 1.25$^{+0.32}_{-0.28}$ & 0.66 & 0.12 & 1.90 & \\
6 & 2.07 & 149.709 & 2.007 & 1.59$^{+0.40}_{-0.34}$ & 0.82 & 0.15 &  & \\
7 & 2.07 & 150.126 & 2.24 & 1.37$^{+0.39}_{-0.33}$ & 0.73 & 0.13 &  & $^e$ \\
8 & 2.20 & 149.609 & 1.774 & 1.37$^{+0.44}_{-0.37}$ & 0.74 & 0.11 &  & \\
9 & 2.21 & 149.843 & 1.807 & 1.34$^{+0.45}_{-0.38}$ & 0.72 & 0.1 &  & \\
10 & 2.23 & 150.326 & 1.89 & 1.33$^{+0.45}_{-0.38}$ & 0.71 & 0.09 & 2.27 & \\
11 & 2.24 & 149.676 & 2.007 & 1.95$^{+0.50}_{-0.43}$ & 0.85 & 0.18 &  & \\
12 & 2.26 & 149.859 & 1.94 & 1.49$^{+0.47}_{-0.39}$ & 0.75 & 0.1 &  & \\
13 & 2.36 & 150.693 & 2.19 & 1.24$^{+0.47}_{-0.39}$ & 0.66 & 0.06 &  & \\
14 & 2.37 & 149.643 & 1.974 & 1.49$^{+0.50}_{-0.42}$ & 0.72 & 0.07 & 2.41 & \\
15 & 2.38 & 150.209 & 1.707 & 1.17$^{+0.46}_{-0.38}$ & 0.63 & 0.05 & 2.42 & \\
16 & 2.39 & 150.476 & 2.657 & 1.25$^{+0.48}_{-0.39}$ & 0.65 & 0.05 &  & \\
17 & 2.42 & 149.809 & 2.124 & 1.91$^{+0.52}_{-0.44}$ & 0.79 & 0.1 & & \\
18 & 2.44 & 149.643 & 2.674 & 1.18$^{+0.46}_{-0.38}$ & 0.63 & 0.04 &  & \\
19 & 2.45 & 150.076 & 2.274 & 1.34$^{+0.49}_{-0.40}$ & 0.67 & 0.05 & & $^f$ \\
20 & 2.48 & 149.576 & 1.957 & 2.08$^{+0.54}_{-0.46}$ & 0.81 & 0.11 & 2.53 & \\
21 & 2.53 & 150.293 & 2.324 & 1.86$^{+0.52}_{-0.44}$ & 0.78 & 0.1 & $_{2.47}^{2.60}$ & $^g$ \\
22 & 2.61 & 149.509 & 1.907 & 1.56$^{+0.52}_{-0.43}$ & 0.71 & 0.08 &  & \\
23 & 2.62 & 150.359 & 2.674 & 1.66$^{+0.53}_{-0.44}$ & 0.74 & 0.09 &  & \\
24 & 2.64 & 150.576 & 2.674 & 1.36$^{+0.50}_{-0.41}$ & 0.67 & 0.06 &  & \\
25 & 2.68 & 150.009 & 2.207 & 1.2$^{+0.50}_{-0.41}$ & 0.63 & 0.05 &  & \\
26 & 2.69 & 150.343 & 2.557 & 1.6$^{+0.54}_{-0.45}$ & 0.71 & 0.08 &  & \\
27 & 2.72 & 149.626 & 2.324 & 1.53$^{+0.55}_{-0.45}$ & 0.7 & 0.07 &  & \\
28 & 2.74 & 149.526 & 1.907 & 2.22$^{+0.63}_{-0.53}$ & 0.82 & 0.14 &  & \\
29 & 2.74 & 150.309 & 2.507 & 1.82$^{+0.59}_{-0.48}$ & 0.76 & 0.1 & 2.77 & \\
30 & 2.77 & 150.009 & 1.974 & 1.35$^{+0.56}_{-0.45}$ & 0.66 & 0.05 &  & \\
31 & 2.81 & 150.343 & 2.64 & 1.93$^{+0.65}_{-0.53}$ & 0.77 & 0.1 & 2.83 & \\
32 & 3.01 & 149.943 & 1.774 & 2.37$^{+0.99}_{-0.77}$ & 0.79 & 0.11 &  & \\
33 & 3.02 & 150.276 & 2.324 & 2.5$^{+1.03}_{-0.79}$ & 0.81 & 0.11 &  & \\
34 & 3.04 & 149.709 & 1.907 & 2.28$^{+1.04}_{-0.79}$ & 0.78 & 0.1 &  & \\
35 & 3.04 & 149.909 & 2.007 & 2.28$^{+1.04}_{-0.79}$ & 0.78 & 0.1 &  & \\
36 & 3.08 & 150.293 & 2.507 & 3.1$^{+1.25}_{-0.96}$ & 0.85 & 0.13 &  & \\
\hline
\end{tabular}
\end{center}
\begin{scriptsize}
$^a$ Galaxy overdensity calculated in a cylindrical window with $r=5'$ and a redshift depth of $0.025(1+z)$.\\
$^b$ Probability of the structure being a proto-cluster with M$_{z=0}>10^{14}$ M$_{\odot}$ given its $\delta_{gal}$ and redshift. $P_{pc}$ corresponds to the sum of the probabilities of all 3 cluster mass bins in Figure 3.\\
$^c$ Probability of the structure being a ``Coma-type'' proto-cluster with M$_{z=0}>10^{15}$ M$_{\odot}$ given its $\delta_{gal}$ and redshift.\\
$^d$ Redshift of its secondary density peak if present.\\
$^e$ Z-FOURGE proto-cluster \citep[z=2.09, ][]{spitler12}.\\
$^f$ HPS proto-cluster ($z=2.44$, Chiang et al. in prep.).\\
$^g$ The signal might be from a line-of-sight filament or multiple structures.\\
\end{scriptsize}
\end{table}

\subsection{Progenitors of Galaxy Clusters in COSMOS/UltraVISTA}

By utilizing the fact that progenitors of structures with higher present day mass will, on average, have a higher $\delta_{gal}$ (Figure 1), we can select proto-clusters in the COSMOS/UltraVISTA field by selecting regions with the highest $\delta_{gal}$. However, not all regions with a high $\delta_{gal}$ will be cluster progenitors due to the significant overlap of the $\delta_{gal}$ distributions between the field and clusters. Also, lower mass structures are much more abundant, which contaminate the high $\delta_{gal}$ regions. This is difficult to see from Figure 1 where the $\delta_{gal}$ distributions are normalized. We have quantified this effect in Figure 3, where we derive the probabilities for a region with a given $\delta_{gal}$ to be non-protocluster (light purple) or a proto-cluster in any of the 3 mass ranges (dark purple). This analysis was done by calculating $\delta_{gal}$ at a large number of random positions in the simulated volume, which is $\sim24$ times larger than the COSMOS field. Any region having $>15\%$ of the galaxies in the cylindrical window which will evolve to a present day cluster was considered to be a ``proto-cluster region''\footnote{This robustly recovers all proto-clusters in the simulation with the level of $z_{phot}$ errors implemented, while it never gives a false positive (except for intrinsically ambiguous cluster-field boundaries). Although non-protocluster regions may  be contaminated by cluster galaxies due to the redshift uncertainties, this never exceeds a few percent.}. Figure 3 quantifies the $\delta_{gal}$ cut required to reach a certain confidence level for cluster progenitor identification. We use $P_{pc}$ and $P_{Coma}$ to specify the probabilities for a structure to be the progenitor of at least a genuine cluster (M$_{z=0}>10^{14}$ M$_{\odot}$) and of a ``Coma-'' like proto-cluster (M$_{z=0}>10^{15}$ M$_{\odot}$), respectively.

We can now select proto-cluster candidates by the following criteria: (1) $\delta_{gal}>\langle\delta_{gal,\ Coma}\rangle$, where $\langle\delta_{gal,\ Coma}\rangle$ is the mean overdensity of ``Coma'' proto-clusters, (2) $P_{pc}\gtrsim0.6$, and (3) visual inspections, requiring that the structures have a fairly smooth overdensity profile on the sky and along the redshift axis, and filament-like structures are excluded. The structures shown in Figure 2 are among the most robust candidates. In total, we obtain 36 proto-cluster candidates in the central $1.2\times1.0$ deg$^2$ of the COSMOS/UltraVISTA field at $1.6<z<3.1$. The candidate list and the derived probabilities are summarized in Table 1. Their estimated positions are based on the local maximums of the smoothed density field, which are accurate to $\sigma_{RA}\sim\sigma_{Dec}\sim1-3$ arcmin and $\sigma_z\sim0.02-0.07$. These structures typically have $P_{pc}\sim70\%$ and $P_{Coma}\sim10\%$. These probabilities were evaluated in exactly the same way as was done for Figure 3, but at the redshift of each individual structure.
%\textbf{(evaluated based on the analysis of Figure 3 but centered at the redshift of each structures saperately)} 

As shown in Figure 2, these candidates show strong galaxy clustering at the scales of $\sim10-20$ arcmin ($\sim15-30$ comoving Mpc). We do not resolve the extension along the line-of-sight due to the $z_{phot}$ errors. In general, structures with redshift separated by $\gg\sigma_z=0.025(1+z)$ are expected to be physically uncorrelated, which should be the case for the candidates listed. However, line-of-sight filaments cannot be completely excluded (e.g., see PC21). Given the $z_{phot}$ uncertainties, a significant fraction of the tracer galaxies of these proto-cluster candidates are likely to be fore- and background interlopers. By examining the regions selected by similar criteria in our matched simulation, we found that $\sim15\%-40\%$ of the photometric redshift galaxy candidates are expected to be true proto-cluster members that will merge into a cluster-scale halo by $z=0$.

Although our technique was specifically designed to find cluster progenitors in the presence of appreciable errors in photometric redshift, until we have spectroscopic confirmation we will consider the targets (Table 1) as (strong) candidates of forming clusters. However, two of our candidates have already been independently confirmed by other surveys, suggesting that our finder is robust. The first candidate, PC07 ($z\sim2.07$), coincides with a structure discovered by \citet{spitler12} in a deep medium-band photometric survey. They found a large number of galaxies in three adjacent clumps of 5 comoving Mpc (diameter), for which spectroscopic follow-up gave a more precise redshift of 2.09. They identified this structure as the progenitor region of a massive cluster. The second candidate, PC19 ($z\sim2.45$), coincides with a large overdensity of Ly$\alpha$ emitters (LAEs) discovered as part of the HETDEX Pilot Survey \citep{adams11}. This particular structure contains 9 bright LAEs in a concentrated peak at $z\sim2.44$, and is also consistent with the properties expected for a forming, massive cluster (Chiang et al. in prep.).

\section{Discussion}

Based on Figure 3, we can estimate the purity of our cluster progenitor finding algorithm by computing the average probability for a structure to be a proto-cluster. For $\delta_{gal}>\langle\delta_{gal,\ Coma}\rangle$, we get a level of purity $\sim70\%$. This purity can be found in another way from our sample of 36 proto-cluster candidates listed in Table 1. The sum of all $P_{pc}$ and $P_{Coma}$ is $\sim26$ and $\sim3$, respectively, which is the number of true cluster progenitors (``Coma-'' type progenitors) expected if a follow-up spectroscopic campaign is performed. The estimated purity from the sample is again $\sim26/36=\langle P_{pc}\rangle\sim70\%$.

We can also estimate the completeness from Figure 1 by calculating the probability for a cluster progenitor to have $\delta_{gal}$ above the threshold we set ($\delta_{gal}>\langle\delta_{gal,\ Coma}\rangle$).  We then get a level of $9\%$, $7\%$, $17\%$, and $50\%$ completeness for all cluster progenitors, ``Fornax-'', ``Virgo-'', and ``Coma-'' type proto-clusters, respectively. These numbers are consistent with our sample of 36 candidates and 26 (3) true proto-clusters (``proto-Comas''): The total comoving volume probed is $\sim2.2\times10^7$ Mpc$^3$ for the central $1.2\times1.0$ deg$^2$ region of the COSMOS field at $1.6<z<3.1$. Scaled from the cluster abundance in the Millennium Simulation, we expect a total $\sim290$ proto-clusters in this volume, which can be further broken down to $\sim240$ ``Fornax-'', $\sim55$ ``Virgo-'', and $\sim5$ ``Coma-'' types, respectively. The number of 26 (3) true proto-clusters (``proto-Coma'') that we found are in good agreement with these numbers when we take into account the completeness. There is a strong trade-off between the purity and completeness. In our case, it is mainly the photometric redshift that blend the $\delta_{gal}$ distributions.

A closer look at Figure 2 and Table 1 shows that there appear multiple structures in the south-west corner of the field at different redshifts. However, they are separated by underdense regions along the line-of-sight with separations $\gg100$ comoving Mpc, suggesting that they are physically uncorrelated. By examining the existing catalog of X-ray structures in this field \citep{Finoguenov07}, we have excluded the possibility that lensing magnification by nearby clusters is responsible for boosting the high redshift number counts.
%in this region \citep{ford13}. We therefore conclude that this region probably contains several unrelated proto-cluster structures along the line-of-sight by chance.

We note that our analysis of the $\sim15$ comoving Mpc scale clustering is solely designed for identifying cluster progenitor structures as a whole and does not necessarily imply the presence of any specific environmental impacts on such scales. However, future studies of the interplay between galaxy properties and environment will benefit from having a large systematic sample of cluster progenitors such as the one presented here.

This work successfully generates a large sample of strong candidates of cluster progenitors in COSMOS, providing a rich set of targets suitable for spectroscopic follow-up that will allow detailed studies of their galaxy properties. Many of our candidates in the central regions of the field may be soon confirmed by spectroscopic redshifts from the zCOSMOS-faint survey (Lilly et al. in prep.), or with VLT/KMOS observations planned in the COSMOS field. It is also possible that these dense proto-cluster regions will show up in planned studies that will use background quasars and bright Lyman break galaxies to perform Lyman-$\alpha$ forest tomographic mapping \citep{KG13}.
Our methods applied to COSMOS can be tuned to search for cluster progenitors with the upcoming spectroscopic/photometric redshift surveys such as HETDEX, Dark Energy Survey, and Subaru Hyper Suprime-Cam and Prime Focus Spectrograph surveys. The much larger volumes and various cosmic epochs probed by these surveys will open up a statistical and multi-dimensional (e.g., mass and redshift) window that will allow a deeper understanding of the (early) formation of galaxies, gas and dark matter in the most extreme cosmic structures.

\begin{acknowledgements}
We thank Edward Robinson and John Silverman for helpful discussions and Adam Muzzin for compiling the COSMOS/UltraVISTA galaxy catalog, which includes datasets from \cite{martin05, capak07, mcCracken12, sanders07}. The Millennium Simulation databases used were constructed as part of the activities of the German Astrophysical Virtual Observatory (GAVO).
%The Millennium Simulation databases used were constructed as part of the activities of the German Astrophysical Virtual Observatory (GAVO).
\end{acknowledgements}

\end{document}